\documentclass[epsf,useAMS,usenatbib,usegraphicx]{mn2e}
\usepackage{epsfig}
\usepackage{aas_macros} 
\usepackage{amsmath}
\usepackage{amssymb}
\usepackage{appendix}
\usepackage{booktabs}
\usepackage{upgreek}
\usepackage{url}
\usepackage[usenames]{color}

\title[Super-Nyquist asteroseismology with TESS]{The potential for super-Nyquist asteroseismology with TESS}

\author[Simon J. Murphy] 
{Simon J. Murphy$^{1,2,\dagger}$\\
$^1$Sydney Institute for Astronomy (SIfA), School of Physics, University of Sydney, NSW 2006, Australia\\
$^2$Stellar Astrophysics Centre, Department of Physics and Astronomy, Aarhus University, DK-8000 Aarhus C, Denmark\\
\\
$^{\dagger}$email: murphy@physics.usyd.edu.au
}
 
\begin{document}

\maketitle 

\begin{abstract}
The perfect 30-min cadence of the full-frame images from the Transiting Exoplanet Survey Satellite (TESS) will impose a hard Nyquist limit of 24\,d$^{-1}$ ($\approx 278$\,$\upmu$Hz). This will be problematic for asteroseismology of stars with oscillation frequencies at or around that Nyquist limit, which will have insurmountable Nyquist ambiguities. TESS does offer some observing slots at shorter cadences, but these will be limited in number and competitive, while the full frame images will be the main data product for many types of variable stars. We show that the Nyquist ambiguities can be alleviated if, when TESS resumes observations after a downlink, integrations are not resumed at perfect cadence with those before the downlink. The time spent idling before integrations are resumed need only be around five minutes for satisfactory results, and observing time can be recouped from the downlink event if the telescope does not wait for a return to perfect cadence before resuming integrations. The importance of imperfect cadence after downlink is discussed in light of phase coverage of transit events.
\end{abstract}


\section{Introduction}

The Transiting Exoplanet Survey Satellite (TESS; \citealt{rickeretal2015}) is due to be launched in 2018. By focussing on brighter targets than the {\it Kepler} mission and by having full sky coverage, TESS presents many new opportunities in the fields of exoplanets and asteroseismology.

TESS will observe in three cadences: full-frame images will be available at a strict 30-min cadence, while individual targets can be observed at \mbox{2-min} and \mbox{20-s} cadences. While some 200\,000 stars will be observed at \mbox{2-min} cadence, these will comprise mostly of cool planet-host candidates. Stars with spectral types earlier than $\sim$F5 will not be observed at the \mbox{2-min} or \mbox{20-s} cadences unless they compete successfully under guest observer or asteroseismic target programmes. In addition, supply of the \mbox{20-s} cadence slots is small, and the demand is expected to be high. Hence, maximising the utility of the 30-min cadence is important.

The sampling situation with TESS is not dissimilar to that of the Kepler Mission. \textit{Kepler} has the capability of observing in a long-cadence (LC) mode of 29.45-min and a short-cadence (SC) mode of 58.85\,s \citep{kochetal2010}. Only 512 SC slots are available, and because of memory and telemetry constraints, not all of these can be used in the successor mission, K2 \citep{howelletal2014}.

The hard Nyquist limit of \textit{Kepler} was overcome due to the unique features of \textit{Kepler}'s heliocentric orbit. The full details were presented by \citet{murphyetal2012b}, but a short summary is given here. In the original mission, a $\pm$200-s R\o{}mer delay existed along the telescope's line of sight towards its field of view between Cygnus and Lyra. As a result, its sampling, which was strictly regular on board the satellite, was periodically modulated by its orbit when those times were converted to barycentric julian date (BJD). The periodic modulation therefore applied to the Nyquist frequency. Nyquist aliases became multiplets in the Fourier transform, when the observations were longer than one \textit{Kepler} orbit of 372.5\,d, while real oscillation peaks remained as singlets. Furthermore, the distribution of power into these multiplets meant that aliases had lower amplitudes than the real peaks. The outcome was that classical pulsators were no longer plagued by Nyquist ambiguity, and the number of stars that could be analysed with confidence increased by nearly two orders of magnitude.

Application of the same methodology to TESS meets problems. TESS is in a geocentric orbit inclined to the ecliptic plane, with a period of just 13.7\,d. The light travel time across its eccentric orbit is only 1.6\,s, which is far shorter than the 1800\,s of one full-frame integration. As such, the modulation to the sampling will not have a detectable effect, and Nyquist ambiguities cannot be resolved in this way. A full explanation is provided in Sect.\,\ref{sec:no_sna}. Of course, the Earth also orbits the Sun, but TESS does not look at any field for one or more Earth orbits except for the continuous viewing zone within 12$^{\circ}$ of the ecliptic pole. Unfortunately, since the R\o{}mer delay is the product of the light travel time across the orbit with the cosine of the latitude of the field, the effect is small, and is applicable to only a tiny fraction of TESS targets.

Fortunately, an alternative solution is available. TESS's step-and-stare operations continue for 27\,d for each field, which equals two orbits. A data downlink occurs at each perigee, during which no observations take place. We show in this paper that the amplitudes of the Nyquist aliases are attenuated if the return to cadence after a downlink is imperfect. This allows the real oscillation frequencies, which are the fundamental data of asteroseismology, to be distinguished from their aliases.

\section{Nyquist aliasing}
\label{sec:nyquist}

Nyquist aliases are peaks in the Fourier transform of the stellar light curve that describe the data equally as well as the true peak. There are an infinite number of Nyquist aliases, and some external physical constraint is required to bound the frequency range of pulsating stars. This could be as fundamental as restricting pulsation time-scales to be longer than the light travel time across the star, but stellar pulsation models can offer much tighter constraints. Still, the frequency ranges of some classes of variable star can straddle one or more Nyquist frequencies of the most commonly used data products, such as the light curves from \textit{Kepler} and those anticipated from TESS.

Nyquist aliases are particularly problematic when the sampling is perfectly regular (see \citet{koen2006,koen2010} for a description of the irregularly sampled case). The Nyquist frequency of regularly sampled data is half the sampling frequency, and is defined as
\begin{equation}
f_{\rm Ny} = f_{\rm S}/2 = 1 / (2 \Delta t),
\end{equation}
where $\Delta t$ is the sampling interval. The time stamp of an observation $t_j$, can therefore be written as
\begin{equation}
t_j = t_0 + j \Delta t
\end{equation}
for an arbitrary (fixed) start time $t_0$.

In order to resolve Nyquist aliases, some modification to the sampling must be made.

\section{Orbital Modulation of the Sampling}
\label{sec:no_sna}

In this section we demonstrate that the super-Nyquist asteroseismology method from \citet{murphyetal2012b}, which utilises the periodic modulation of the sampling due to the orbit of the satellite, will not be viable for resolving Nyquist ambiguity with TESS. The full mathematical foundation for this analysis was presented by \citet{murphyetal2012b}, and so the interested reader is referred there. Given the null result here, we restrict this discussion to only the most relevant points.

The super-Nyquist asteroseismology method as applied to \textit{Kepler} relies on the substantial modification of the time-stamps due to orbital motion. Since \textit{Kepler} orbits the Sun with a semi-major axis, $a$, of 500 light-seconds, light arrives at the telescope early or late, compared to the solar system barycentre. Thus, when the observation times are corrected to the arrival time at the solar system barycentre, a periodic modulation is imprinted. The time delay was described by \citet[][Eq. 11 there]{murphyetal2012b} as
\begin{equation}
\delta t (t) = \frac{a}{c} \cos \beta \cos \{ \lambda_{\odot}(t) - \lambda \},
\label{eq:delta_t}
\end{equation}
where $\lambda$ and $\beta$ are the ecliptic longitude and latitude of the \textit{Kepler} field. The $\cos \beta$ term reduces the effective R\o{}mer delay,
\begin{equation}
\tau = \frac{a}{c} \cos \beta
\label{eq:tau}
\end{equation}
to $\pm 200$\,s, which is still a substantial fraction of the $\sim$1770\,s sampling interval of \textit{Kepler} long-cadence mode.

\citet{murphyetal2012b} showed that the periodic modulation of the sampling causes aliases in the Fourier transform to become multiplets, while the real angular oscillation frequency, $\omega_R \equiv 2 \uppi f_{\rm R}$, remains a single peak. The central components of each multiplet are found at the anticipated alias frequencies, that is, at $n\omega_S \pm \omega_R$, where $\omega_S$ is the sampling frequency and $n$ is an integer, while the additional components of the multiplets are separated from the central component by integer multiples of the angular orbital frequency ($\Omega_{\rm orb} \equiv 2 \uppi / P_{\rm orb}$), that is, at $n\omega_S \pm \omega_R \pm m\Omega_{\rm orb}$. The amplitudes of the $m$-th components of the multiplet with respect to the central peak are determined from the ratio of the Bessel coefficients $J_m(n\omega_S \tau) / J_0(n\omega_S \tau)$.

Application of Eq.\,\ref{eq:tau} to TESS yields a maximum R\o{}mer delay of $\tau \simeq0.8$\,s, if the eccentricity of the orbit is neglected. Since TESS's orbit is inclined to the ecliptic, the relevant angle in Eq.\,\ref{eq:tau} is no longer the ecliptic latitude, but the angle between the satellite's orbital plane and the target, which we call $\beta'$, and which still has the minimum value of $0^{\circ}$. The TESS full-frame images will have a cadence approximately equal to the \textit{Kepler} long-cadence mode, thus the amplitudes of the non-central components of the alias multiplets will be reduced by a factor of $\tau_{\rm Kepler} / \tau_{\rm TESS} = 200/0.8 = 250$. Relevant parameters of the TESS 30-min sampling are given in Table\:\ref{tab:TESS_LC_parameters}.

\begin{table}
\centering
\caption{Parameters for TESS 30-min sampling.}
\begin{tabular}{ccl}
\toprule
Parameter & Value & Units\\
\midrule
$\tau$ & $9.26 \times 10^{-6}$ & d\\
$\Delta t$ & $2.08 \times 10^{-2}$ & d\\
$\omega_S$ & $3.02 \times 10^{+2}$ & rad\,d$^{-1}$\\
$\Omega_{\rm orb}$ & $4.59 \times 10^{-2}$ & rad\,d$^{-1}$\\
\bottomrule
\end{tabular}
\label{tab:TESS_LC_parameters}
\end{table}

\begin{figure}
\begin{center}
\includegraphics[width=0.48\textwidth]{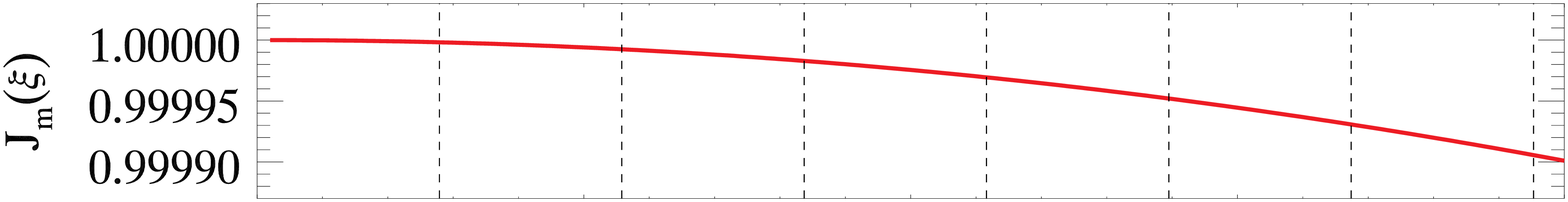}
\includegraphics[width=0.48\textwidth]{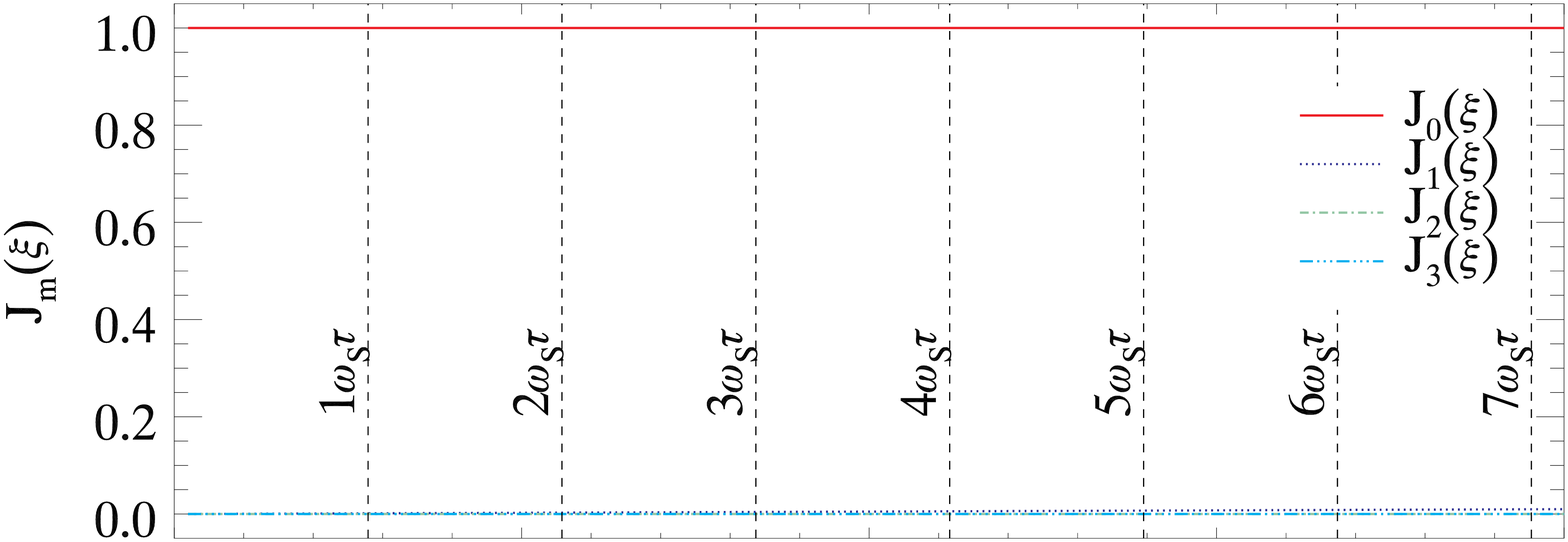}
\includegraphics[width=0.48\textwidth]{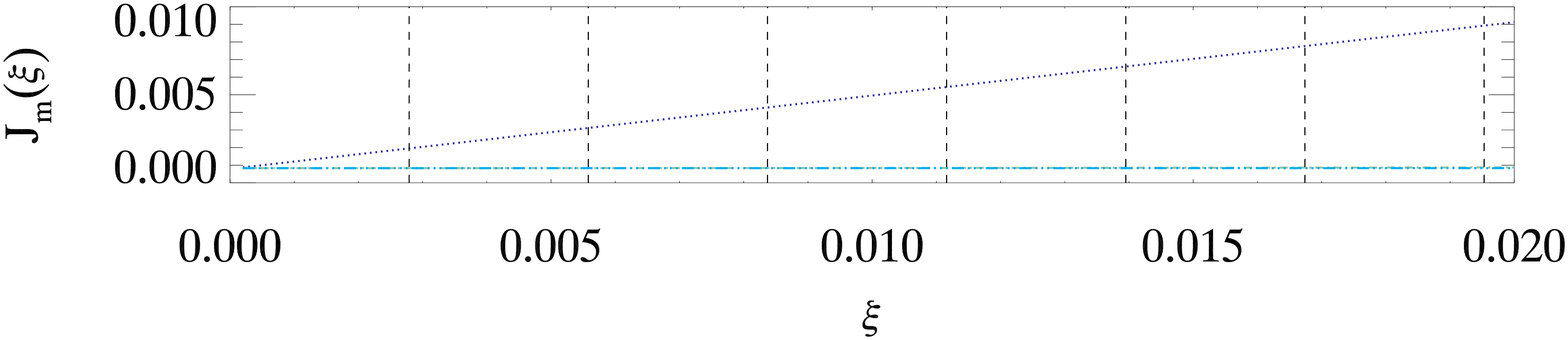}
\caption{The Bessel coefficients $J_m(\xi)$ with $m = [0\,..\,3]$ for $\xi = [0\,..\,0.02]$. The vertical lines show $\xi = n \omega_{\rm S} \tau$ with $n = [0\,..\,7]$, at which the window spectrum has sharp but small multiplet peaks. The centre panel shows all Bessel coefficients, while the top and bottom panels zoom in to show the variation of $J_0(\xi)$ and $J_{1..3}(\xi)$, respectively. The highest values of the coefficients of $J_{2}(\xi)$ and $J_{3}(\xi)$ in this range are $5.0\times10^{-5}$ and $1.6\times10^{-7}$, respectively.}
\label{fig:bessel}
\end{center}
\end{figure}

By computing the Bessel coefficients $J_m(n\omega_S \tau)$ for \mbox{$n = [0\,..\,7]$} and \mbox{$m = [0\,..\,3]$} we can evaluate the multiplet structure for the first $2n+1=15$ Nyquist aliases, and determine whether these multiplets will be detectable in the Fourier transform of typical TESS observations of classical pulsators. The small value of $\tau$ for the TESS orbit causes the Bessel coefficients to change very slowly with $\xi \equiv \omega \tau$, as indicated in Fig.\,\ref{fig:bessel}. The implications are that the Nyquist aliases are self-similar and cannot be used to resolve Nyquist ambiguity, even in the most favourable case of a target in the orbital plane (i.e. with $\beta' = 0$). The amplitudes of the first sidelobes of the multiplets are tiny, as given in Table\:\ref{tab:multiplets}, and more distant sidelobes (higher $m$) are completely negligible (see Fig.\,\ref{fig:bessel} caption). Although the first sidelobes do become detectable in higher order aliases (higher $n$), the self-similarity of neighbouring aliases does not permit distinction between them, and the real peak cannot be identified.

\begin{table}
\centering
\caption{Amplitude ratios of the first sidelobes of an alias multiplet ($m=1$) to the central component ($m=0$), as a function of the alias identification (of the form $n\omega_{\rm S}\pm\omega_{\rm R}$), for TESS 30-min sampling. The amplitude ratios for negative values of $n$ are identical to their positive values.}
\begin{tabular}{ccl}
\toprule
$\lvert n \rvert$ & Amp.($m=1$) / Amp.($m=0$) \\
\midrule
0 & 0.000000 \\
1 & 0.001350 \\
2 & 0.002750 \\
3 & 0.004150 \\
4 & 0.005550 \\
5 & 0.006950 \\
6 & 0.008350 \\
7 & 0.009750 \\
\bottomrule
\end{tabular}
\label{tab:multiplets}
\end{table}

An approximate anticipated visibility of the sidelobes observed by TESS can be calculated. This example uses $\delta$\,Sct stars, which are classical pulsators with low-order p\:modes and periods in the range of $\sim15$\,min to 5\,hr, but similar examples can be constructed for other classes of pulsating stars with coherent oscillations. The first pair of Nyquist aliases for a star in the ecliptic plane will have sidelobes with maximum amplitudes of [$J_1(\omega_S \tau)/J_0(\omega_S \tau)=$] 0.135\:per\:cent of that of the central component. Given an estimated standard deviation of relative flux of 200\,ppm/hr$^{1/2}$ for a 10th magnitude star \citep{sullivanetal2015}, and fields of 27.4\,d = 657.6\,hr (neglecting the central downlink gap), then an optimistic noise floor of 8\,ppm will apply. While there are some \textit{Kepler} $\delta$\,Sct stars with amplitudes of tens of mmag (e.g. KIC\,9700322: \citealt{bregeretal2011}, \citealt{guzik&breger2011}; KIC11754974: \citealt{murphyetal2013a}), \textit{Kepler} has shown that typical $\delta$\,Sct stars have amplitudes on the order of 1\,mmag \citep{balona&dziembowski2011}, or $\sim$1\,ppt of relative flux. The TESS passband is redder than that of \textit{Kepler}, so oscillation amplitudes will be lower, but we assume amplitudes of 1\,ppt for ease of calculation. Then the maximum sidelobe amplitudes are on the order of 1\,ppm for the first Nyquist alias, and only around 7\,ppm for the fifth Nyquist alias (that is, at $\xi = 5 \omega_{\rm S} \tau$; see Table\:\ref{tab:multiplets}). Brighter stars have a lower noise floor, estimated to be as low as 60\,ppm/hr$^{1/2}$ \citep{sullivanetal2015}. However, in reality the spectral window of the many different Fourier peaks in multi-periodic oscillators will contribute extensively to the noise level, and such sidelobes will remain undetectable.

\begin{figure}
\begin{center}
\includegraphics[width=0.48\textwidth]{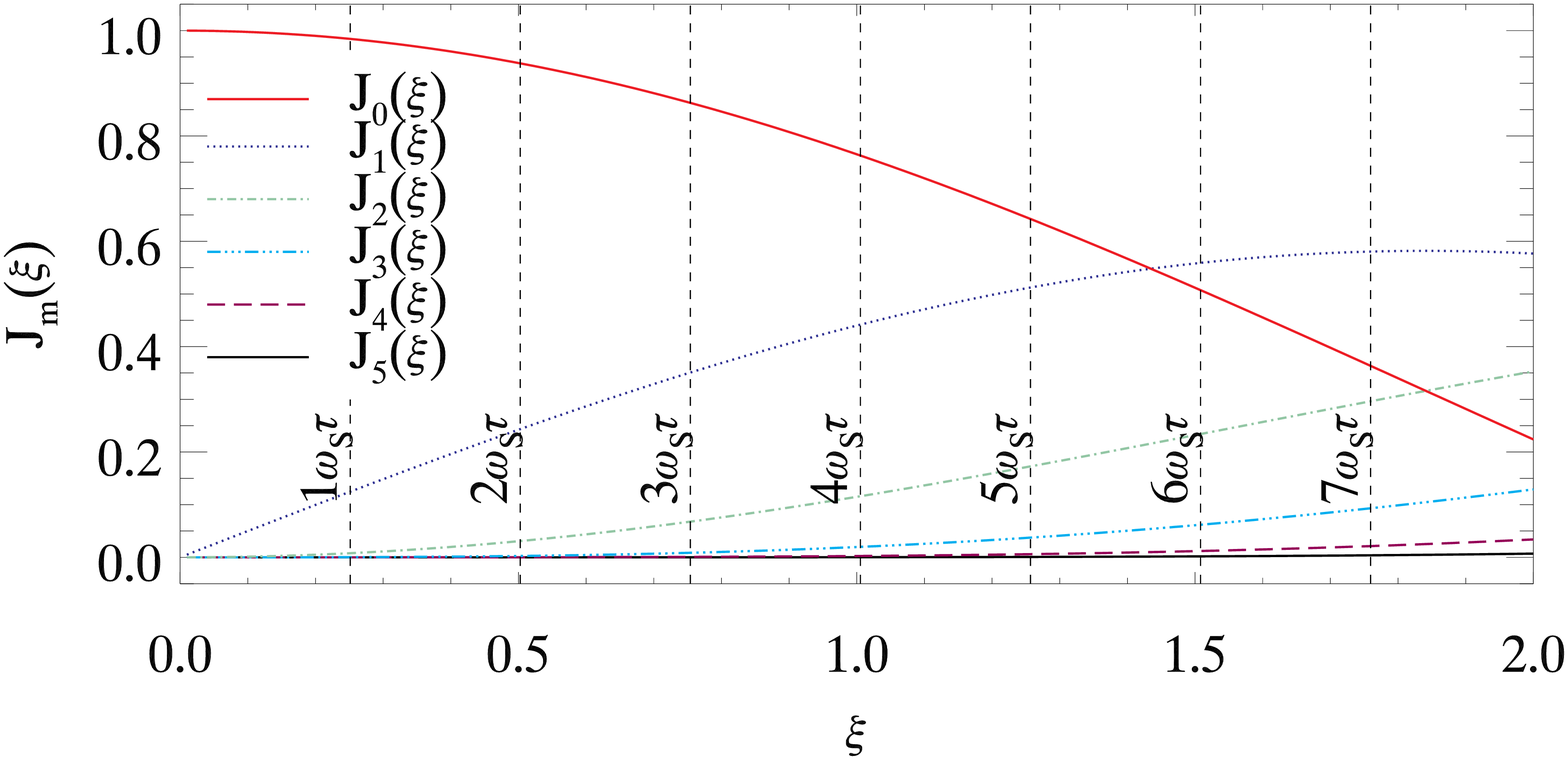}
\caption{The Bessel coefficients $J_m(\xi)$ with $m = [0\,..\,5]$ for $\xi = [0\,..\,2.0]$ for the TESS 20-s cadence where $\omega_{\rm S} = 2.71\times10^{4}$\,rad\,d$^{-1}$. The vertical lines show $\xi = n \omega_{\rm S} \tau$ with $n = [0\,..\,7]$, at which the window spectrum has sharp multiplet peaks with significant sidelobe amplitudes, which could be used to distinguish real peaks from their Nyquist aliases.}
\label{fig:bessel_20s}
\end{center}
\end{figure}

Thus far, only the 30-min cadence of TESS has been discussed. The Bessel coefficients depend on the sampling frequency, therefore the sidelobes of the alias multiplets will be much higher in the shorter cadence modes (Fig.\,\ref{fig:bessel_20s}). However, these modes also have higher Nyquist frequencies (4320\,d$^{-1}$ = 50\,mHz for the 20-s cadence), and so there will be no Nyquist ambiguity in need of resolution.

In the next section we present a promising alternative approach that will enable Nyquist ambiguity to be resolved in all observing modes, and that will additionally benefit the main mission in the discovery of exoplanets.

\section{Offsetting the sampling}

\begin{figure*}
\begin{center}
\includegraphics[width=0.85\textwidth]{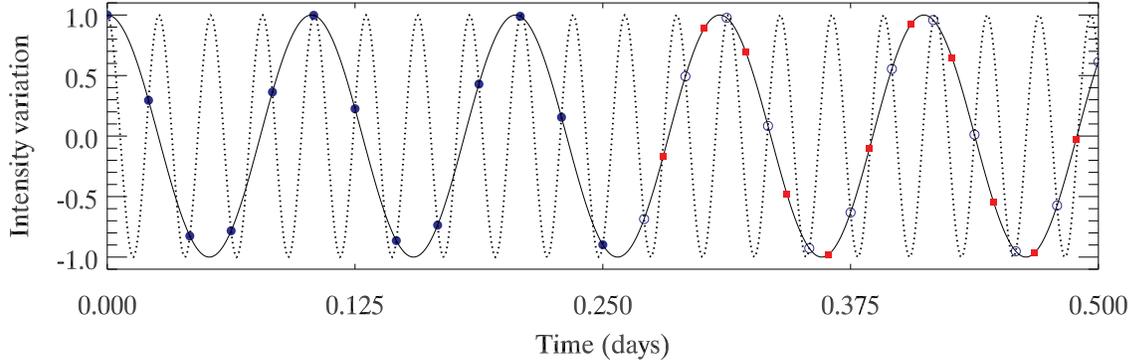}
\caption{Simulated TESS data with 30-min sampling are shown as blue circles. Half-way through the observations an offset is introduced, altering the sampling to the red squares, after which the cadence remains a strict 30\,min. The sampled oscillation has a frequency of 9.71\,d$^{-1}$ (solid line), which has a corresponding Nyquist alias at 38.29\,d$^{-1}$ (dashed line). If the cadence had remained strictly regular at 30\,min then the real peak and its alias would have fitted the data equally well, but due to the offset only the real peak remains a good fit.}
\label{fig:nyquist_demo}
\end{center}
\end{figure*}

If the sampling interval is irregular, or an offset is introduced, Nyquist aliasing can be alleviated, as illustrated in Fig.\,\ref{fig:nyquist_demo}. The data shown there are simulated TESS data with 30-min sampling, for a monoperiodic star oscillating at 9.71\,d$^{-1}$ (=112.4\,$\upmu$Hz).\footnote{The oscillation frequency was chosen to not have a simple integer relationship with the sampling frequency.} Half way through the data set, a single offset of half a cadence (15\,min) is introduced, and thereafter the cadence remains fixed at 30\,min. That is, the time stamps in the second half of the data set are now expressed as
\begin{equation}
t_j = t_0 + (j + \epsilon) \Delta t,
\end{equation}
where $\epsilon$ is the introduced offset that in this case is equal to 0.5. The offset causes the first Nyquist alias to be in anti-phase with the observations, allowing it to be easily distinguished from the real peak.

\begin{figure}
\begin{center}
\includegraphics[width=0.45\textwidth]{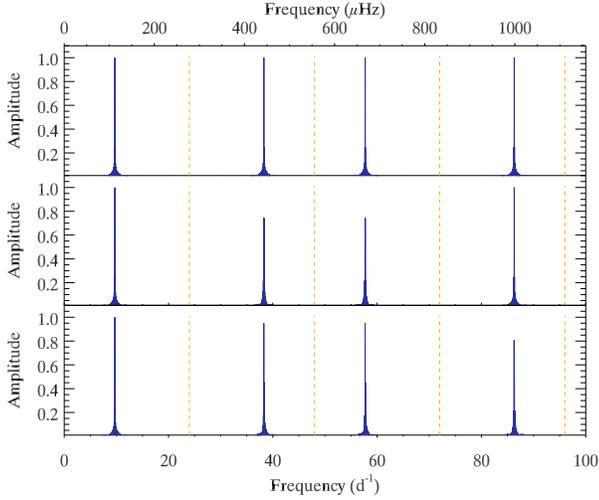}
\caption{Fourier transforms for three different values of the offset parameter $\epsilon$: top = 0.0, middle = 0.5, bottom = 0.213. Dashed orange lines are drawn at integer multiples of the Nyquist frequency, $f_{\rm Ny} = 24.0$\,d$^{-1}$. The input frequency, $f_{\rm in}$, is 9.71\,d$^{-1}$.}
\label{fig:sampling_FTs}
\end{center}
\end{figure}

The anticipated TESS data are of longer duration than those shown in Fig.\,\ref{fig:nyquist_demo}, and the proposed sampling offset would be introduced after each downlink, which are of ``no more than 16\,h'' \citep{rickeretal2015}. Therefore more realistic 27-d simulations were created with a central gap during which an offset is introduced. The gap was set at 0.8\,d, corresponding to a pessimistic 19-h duration, but the exact length is unimportant. The Fourier transform of those simulations for a monoperiodic star oscillating at an input frequency of $f_{\rm in} = 9.71$\,d$^{-1}$ are shown in Fig.\,\ref{fig:sampling_FTs}, for values of $\epsilon = 0$, 0.5 and 0.213. The latter value was chosen to be approximately intermediate between the in-phase ($\epsilon = 0 \equiv 1$) and anti-phase ($\epsilon = 0.5$) cases, but not have a simple relationship with either 0.5 or 1.

\begin{figure}
\begin{center}
\includegraphics[width=0.49\textwidth]{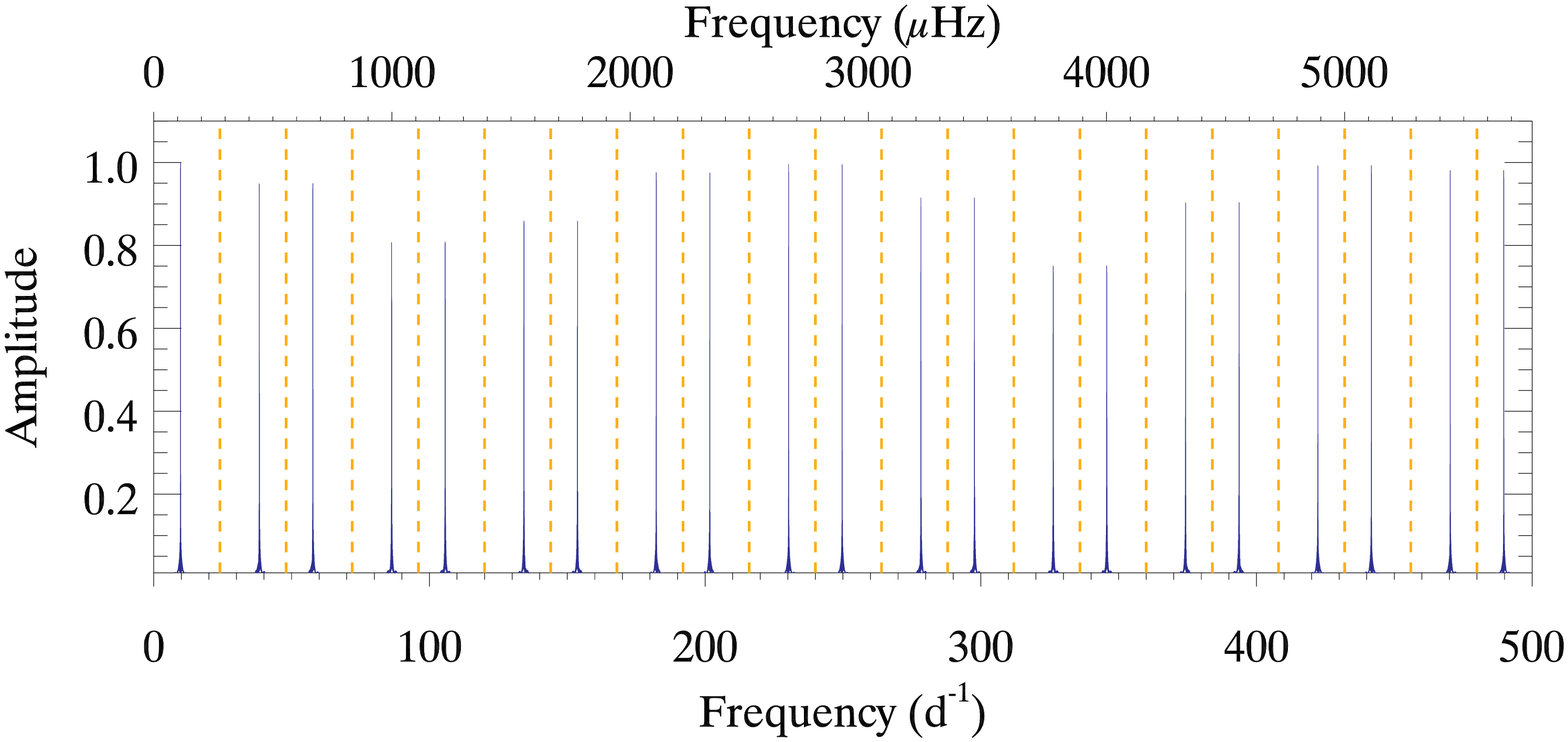}
\caption{Fourier transform of 27\,d of simulated TESS data with an offset of $\epsilon=0.213$. Dashed orange lines are drawn at integer multiples of the Nyquist frequency (=24\,d$^{-1}$). The input frequency, $f_{\rm in} = 9.71$\,d$^{-1}$, has the highest amplitude up to a frequency of 230\,d$^{-1}$, where a pair of aliases have similar amplitudes.}
\label{fig:high_freq}
\end{center}
\end{figure}

It can be seen in the top panel of Fig.\,\ref{fig:sampling_FTs} that if no offset is introduced, the Nyquist aliases all have equal amplitudes and the identification of the real peak is totally ambiguous. However, introduction of an offset of $\epsilon = 0.5$ causes the Nyquist aliases at $f_{\rm S} \pm f_{\rm in}$ to have amplitudes lower than the input peak. In this way, the Nyquist ambiguity has been resolved and the input frequency can be recovered. But notice the alias peak near 87\,d$^{-1}$ ($\approx 1000$\,$\upmu$Hz), whose amplitude is still equal to that of the input frequency. This alias can only be discarded if we have an external physical constraint with which to select the real peak. In practice this tends to be the case, because here the real and ambiguous alias peak are an order of magnitude different in frequency. However, the bottom panel of Fig.\,\ref{fig:sampling_FTs} shows that there is no need; an offset that is not a simple fraction of 0.5 or 1.0 will produce a series of aliases whose amplitudes are all lower than that of the input frequency, out to much higher frequency in the Fourier transform (see Fig.\,\ref{fig:high_freq}). The Nyquist ambiguity can then be considered completely resolved. The reason is twofold: (1) any astrophysical target for which ambiguity at such high frequency could be an issue would presumably have been proposed for the shorter cadence slots anyway; and (2), oscillation frequencies so much higher than the sampling frequency will have severe amplitude attenuation (a factor $\sim$10 reduction at 230\,d$^{-1}$) and will most likely be unsuitable for study (see \S\,\ref{ssec:amplitude_reduction}).

\section{Super-Nyquist asteroseismology}

\begin{figure}
\begin{center}
\includegraphics[width=0.49\textwidth]{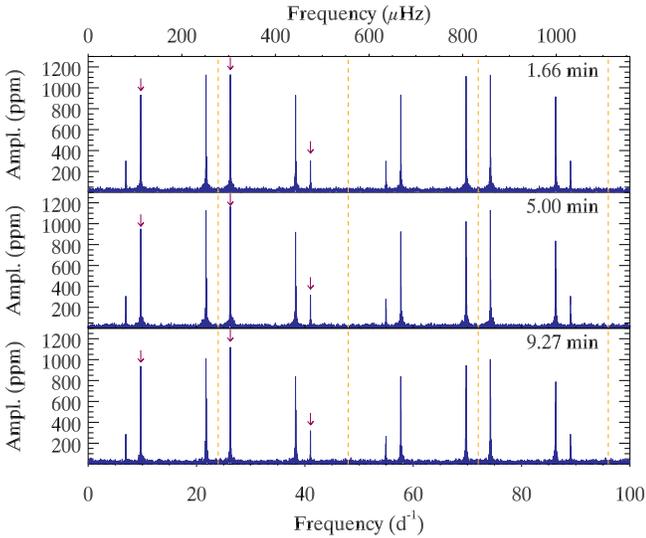}
\caption{Fourier transform of 27\,d of simulated TESS data with different values of the offset, $\epsilon$. Dashed orange lines are drawn at integer multiples of the Nyquist frequency (=24\,d$^{-1}$). The input frequencies, identified with arrows, are 9.710, 26.240 and 41.033\,d$^{-1}$. They are distinguishable from their aliases by their higher amplitudes. The top, middle and bottom panels have offsets of $1.66, 5.00$ and $9.27$\,min, respectively, which are equal to fractional cadences of $\epsilon = 0.0556$, 0.1667 and 0.3090.}
\label{fig:multiperiodic}
\end{center}
\end{figure}

With the introduction of a suitable offset, real peaks can be distinguished from aliases even when those real peaks exceed the Nyquist frequency. In Fig.\,\ref{fig:multiperiodic} the general case is shown, where the star is multiperiodic and the input oscillations are distributed above and below the Nyquist frequency, at 9.710, 26.240 and 41.033\,d$^{-1}$. These input oscillations were given random phases. Noise was added to the simulated data at a level anticipated for a 13th magnitude star.

The input oscillations are entirely recoverable by selecting the peak with the highest amplitude from the set of Nyquist ambiguities for each mode. Even for this 13th magnitude star, the amplitude differences in the bottom panel of Fig.\,\ref{fig:multiperiodic} are significant to 10\,$\sigma$. Therefore, random noise spikes will not push an alias to an amplitude higher than the real peak.

Close inspection of Fig.\,\ref{fig:multiperiodic} provides insight on the optimal offset for the purpose of distinguishing aliases. While amplitude differences are still significant for the 1.66 and 5.00-min offsets, it is clear that the 9.27-min offset is more favourable. The reasons are as follows. For the 1.66-min offset, the phase difference between the true peak and its alias has had little time to accumulate. An offset of 5.00\,min, i.e 1/6 of the 30-min cadence, trebles the available accumulation time, but it is preferable to avoid simple fractions of a cadence. The 9.27-min offset was chosen to not be a simple fraction and to be roughly intermediate between the fractional cadences of 0.0 and 0.5 discussed in the previous section. The amplitude differences are much larger as a result, as is shown in the next subsection.

\subsection{Amplitude reduction of real and alias peaks}
\label{ssec:amplitude_reduction}

\begin{figure}
\begin{center}
\includegraphics[width=0.45\textwidth]{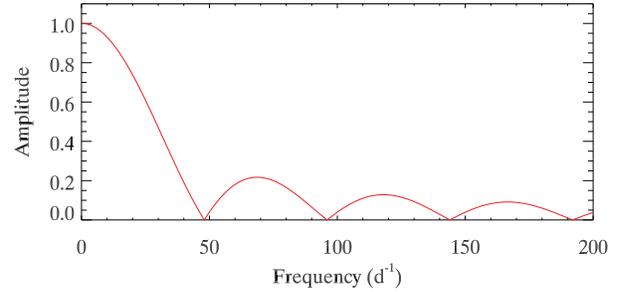}
\caption{Measured amplitude in arbitrary units as a function of frequency, for TESS data with a sampling frequency of 48\,d$^{-1}$. The input amplitude was 1.0 on this scale.}
\label{fig:amp_red}
\end{center}
\end{figure}

The effect of sampling on oscillation amplitudes as applied to \textit{Kepler} data was described in detail by \citet{murphy2014}. The amplitude reduction due to undersampling is a sinc function \citep{chaplinetal2011,murphy2012a}:
\begin{equation}
\mathcal{A}_{\rm obs} = \mathcal{A}_{\rm in}\:{\rm sinc} (\uppi/x),
\label{eq:amp_red}
\end{equation} 
where $x$ is the number of data points per oscillation cycle. This amplitude reduction is illustrated in Fig.\,\ref{fig:amp_red}.

\begin{figure*}
\begin{center}
\includegraphics[width=0.90\textwidth]{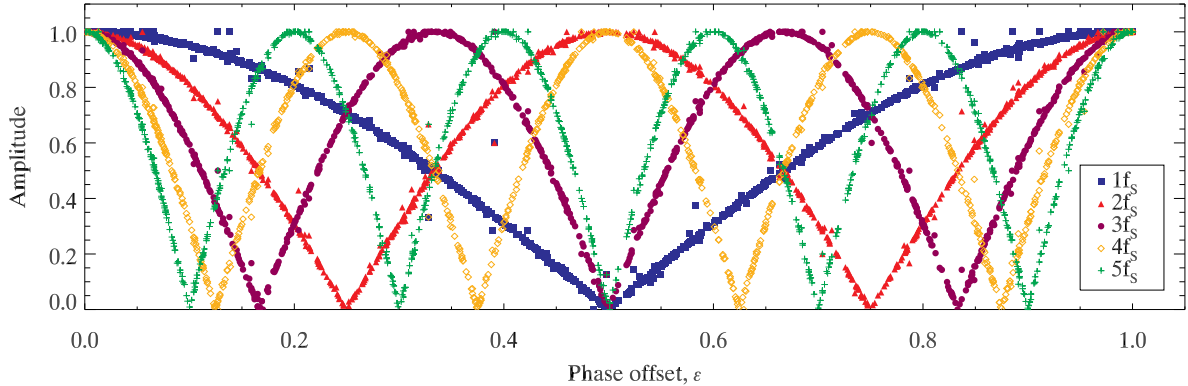}
\caption{Amplitude ratios of alias peaks at $n f_{\rm S} \pm f_{\rm R}$ to real peaks at $f_{\rm R}$, for integer $n = [1\,..\,5]$. Aliases have amplitudes below their corresponding real peak, as long as $\epsilon$ is not a simple fraction. Noisy data contribute some scatter. Based on 1000 simulated light curves.}
\label{fig:amplitudes_vs_epsilon}
\end{center}
\end{figure*}

In this work, the amplitudes of greatest interest are those of the aliases with respect to the real peak. The aliases come in pairs of the same amplitude, at frequencies of $n f_{\rm S} \pm f_{\rm R}$, where $f_{\rm S}$ and $f_{\rm R}$ are the sampling and `real' oscillation frequency, respectively, and $n$ is an integer coefficient. In Fig.\,\ref{fig:amplitudes_vs_epsilon}, the ratios of the measured alias amplitudes, $\mathcal{A}(n f_{\rm S} \pm f_{\rm R})$, to the measured amplitude of the real peak, $\mathcal{A}(f_{\rm R})$, are shown as a function of the sampling offset, $\epsilon$. The ratios were determined by simulating 1000 artificial TESS light curves of 27\,d, with added noise, a central gap of 0.8\,d, and a (uniformly) randomly generated $\epsilon$ for each light curve. The light curves were of a monoperiodic oscillator, each with a different oscillation frequency drawn from a uniform random distribution between 0.0 and 100.0\,d$^{-1}$. The outcome is independent of those oscillation frequencies; the only important factors are $n$ and $\epsilon$. The functional form of the amplitudes in Fig.\,\ref{fig:amplitudes_vs_epsilon} is
\begin{equation}
\frac{\mathcal{A}(n f_{\rm S} \pm f_{\rm R})}{\mathcal{A}(f_{\rm R})} = \lvert \cos (n \uppi \epsilon) \rvert.
\label{eq:functional_form}
\end{equation}

It is easily seen from Eq.\,\ref{eq:functional_form} and Fig.\,\ref{fig:amplitudes_vs_epsilon} that aliases at \mbox{$n f_{\rm S} \pm f_{\rm R}$} will have the same amplitudes as the real peak at $f_{\rm R}$ if \mbox{$\epsilon = i / n$}, where $i$ is an integer. Since both $i$ and $n$ are integers, simple fractions for $\epsilon$ are best avoided so that we can distinguish aliases from real peaks. Ideally we would choose $\epsilon$ to be irrational, e.g. as $1/\uppi$, but in practice $\epsilon$ is best chosen randomly. Specifically, TESS should return to observations as soon as it is ready, and should not wait in an attempt to return to perfect cadence. The most important distinction to be made is between the first alias pair ($n = 1$) and the real peak, since these are closest in frequency. Hence if $\epsilon$ were carefully selected, values nearer to (but not equal to) 0.5 would be preferable to those nearer to 0.0 ($\equiv 1.0$).

\subsection{Other cadences}

Carefully planned offsets could be inserted into the shorter cadence observations to achieve the same effect. The \mbox{2-min} cadence is only six times longer than the \mbox{20-s} cadence, and so the introduction of a suitable offset to that \mbox{2-min} cadence could push its effective Nyquist frequency to roughly the same value as that of the \mbox{20-s} cadence. This does not relieve the amplitude reduction effect due to undersampling (Eq.\,\ref{eq:amp_red}), so it will remain the case that \mbox{20-s} sampling is warranted for high-frequency, low-amplitude oscillators such as white dwarfs. It is noteworthy that if a random offset is introduced for the purposes of the 30-min full-frame images, then the shorter cadences will automatically benefit from the same offset, though the fractional cadence of that offset will of course differ according to the length of the cadence considered.


Other photometric datasets can supplement TESS and \textit{Kepler} data to resolve Nyquist ambiguities. For instance, the SuperWASP \citep{pollaccoetal2006} and KELT \citep{pepperetal2007} data have random sampling, and oscillations can be detected at frequencies much higher than the mean sampling frequency. This is exemplified by the discovery of high-frequency oscillators in SuperWASP data \citep{holdsworthetal2014}, which included rapidly oscillating Ap stars and the fastest $\delta$\,Sct pulsator known.

\subsection{Application to the continuous viewing zone}

\begin{figure*}
\begin{center}
\includegraphics[width=0.85\textwidth]{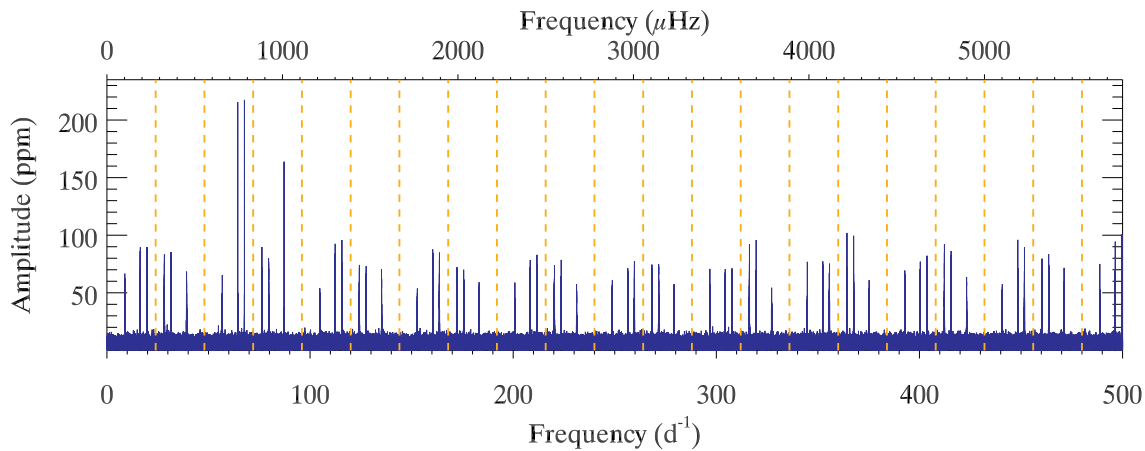}
\caption{Fourier transform of 351\,d of simulated TESS data, representative of the continuous viewing zone. The value of the offset, $\epsilon$, is randomised every orbit. Dashed orange lines are drawn at integer multiples of the Nyquist frequency (=24\,d$^{-1}$). The input frequencies at 64.4, 67.7 and 87.2\,d$^{-1}$ are completely recovered. The noise, corresponding to the prediction for a 15th magnitude star, has no effect on the ability to apply the method.}
\label{fig:continuous}
\end{center}
\end{figure*}

Some parts of the sky will be observed for more than 27\,d. The TESS continuous viewing zones at the ecliptic poles will be observed for 351\,d each. If the return to cadence is imperfect every time, that is, some random value of $\epsilon$ is implemented, then super-Nyquist asteroseismology is made easier. An example is provided in Fig.\,\ref{fig:continuous}. The amplitude of the real peaks are at least twice those of the aliases, making the real peaks easy to identify, even when they have frequencies greater than the sampling frequency. Those amplitudes are still heavily reduced because the exposure time is longer than the pulsation period. The actual input amplitudes for the real peaks at 64.4, 67.7 and 87.2\,d$^{-1}$ were 1000, 1000 and 2000\,ppm, respectively, and so the amplitude reduction shown in Fig.\,\ref{fig:amp_red} is observed.

\section{Conclusions}

The short light-travel time across the TESS orbit causes only a small R\o{}mer delay. The resulting modulation of the sampling interval over the orbital period is insufficient to produce detectable multiplets out of the Nyquist aliases in 30-min TESS full-frame images.

However, the Nyquist limits of all TESS cadences can be raised by introducing a sampling offset while the spacecraft performs a data downlink. Providing that the time of return to observations after downlink is not an integer number of cadences since the last observation, Nyquist ambiguities can be resolved. It was shown that an offset from perfect cadence with no simple integer relationship to a full cadence is preferable. Observing time can be recouped from the downlink event by implementing this proposed offset, because for the application to the full-frame images, the telescope need not wait for the observing gap to reach an integer multiple of 30\,min in duration.

Application to the continuous viewing zone yields even more straightforward identification of the real peaks, because the amplitudes of the aliases are more strongly reduced as a result of several introduced sampling offsets.

Sampling offsets are also important to the main goal of the mission. With the huge number of transiting events that TESS will detect, it is inevitable that some celestial orbital periods will be an integer (or near-integer) multiple of the sampling interval, which will lead to poor transit phase coverage. Offsetting the sampling after downlink will alleviate this and open up additional opportunity for transit timing variations TTVs for the very short-period planets, or for planets in the continuous viewing zone.

The major consequence of this work is that thousands more stars will be opened up for asteroseismic study.

\section*{Acknowledgements}
This research was supported by the Australian Research Council. Funding for the Stellar Astrophysics Centre is provided by the Danish National Research Foundation (grant agreement no.: DNRF106). The research is supported by the ASTERISK project (ASTERoseismic Investigations with SONG and Kepler) funded by the European Research Council (grant agreement no.: 267864).\\

\bibliography{sjm_bibliography} 
\end{document}